\documentclass[aps,prd,superscriptaddress]{revtex4}
\begin{document}

\title{Local correlations in a strongly interacting 1D Bose gas}

\author{D.M.~Gangardt}
\affiliation{\mbox{Laboratoire Kastler-Brossel, Ecole Normale Sup\'erieure, 
24 rue Lhomond, F-75231, Cedex 05, Paris, France}}
\author{G.V.~Shlyapnikov}
\affiliation{\mbox{Laboratoire Kastler-Brossel, Ecole Normale Sup\'erieure, 
24 rue Lhomond, F-75231,
  Cedex 05, Paris, France}}
\affiliation{
\mbox{FOM Institute for Atomic and Molecular Physics, Kruislaan 407,
   1098 SJ Amsterdam, The Netherlands}}

\affiliation{Russian Research Center, Kurchatov Institute, Kurchatov
  Square, 123182 Moscow, Russia}

\date{\today}

\begin{abstract}
We develop an analytical method for calculating local correlations in strongly
interacting 1D Bose gases, based on the exactly solvable Lieb-Liniger model.
The results are obtained at zero and finite temperatures. They describe the
interaction-induced reduction of local many-body correlation functions and can
be used for achieving and identifying the strong-coupling Tonks-Girardeau
regime in experiments with cold Bose gases in the 1D regime. 
\end{abstract}

\maketitle

\section{Introduction}
Recently, one-dimensional (1D) Bose gases have been created in long
cylindrical traps by tightly confining the transverse motion of particles to
zero-point oscillations \cite{ExpGr,ExpGo,ExpSc}. In the present stage, one of
the main goals is to achieve the strong-coupling Tonks-Girardeau (TG) regime
\cite{Girardeau60-65,LiebLiniger1963}, where due to repulsion between
particles the 1D Bose gas acquires fermionic properties.  These studies
have revived an interest in dynamical and correlation properties of 1D gases.
In most cases, correlations and dynamics of trapped atomic gases in the 1D
regime can be investigated \cite{TPSW,TGW,TDLO,TMS,TOS,Gangardt,Karen} 
by using the  Lieb-Liniger model
\cite{LiebLiniger1963}, which assumes that particles interact via a
delta-function repulsive potential. Of particular importance are local 3-body
and 2-body correlations \cite{Gangardt,Karen} as they govern the rates of
inelastic processes, such as 3-body recombination and photoassociation in pair
collisions. The measurement of these rates provides a way to identify
the strong-coupling Tonks-Girardeau regime, and the intrinsic decay process of
recombination in 3-body collisions is crucial for the stability of the gas.

The Lieb-Liniger model for a uniform  system of $N$ bosons  
belongs to the class of integrable models of statistical physics
\cite{Exact1D}. The ground state and spectrum of elementary excitations
for this model were found by Lieb and Liniger 
\cite{LiebLiniger1963} and a theory for finding thermodynamic functions 
at finite temperatures
was constructed  by C.N. Yang and C.P. Yang \cite{Yang1969}. These
quantities are determined by relatively simple integral
equations for the distribution of quantum numbers (quasi-momenta
or rapidities). On the other hand,  the problem of correlation 
properties is far from being completely resolved, except for some
correlation functions in the  limiting cases of weak and strong interactions.
For example, the case of infinitely strong
interactions is to a certain extent  equivalent to that of
free fermions and the interactions play the role of Pauli
principle \cite{Girardeau60-65}. In this limit, any
correlation function of the density is given by the corresponding
expression for fermions, and the one-particle density matrix
is  the  determinant of a certain integral operator \cite{Lenard,Vaidya}. 
The expressions for the one-body and two-body correlations for an arbitrary
interaction strength were obtained by using the Inverse
Scattering Method \cite{KorepinBook}. However, closed analytical results can
be found only as perturbative expansions in the limiting cases of weak and
strong interactions \cite{Jimbo,Korepin,Creamer}. 

In this paper, we calculate local many-body correlations, that is the
expectation value of a product of $2m$ field operators ($m=2,3,\ldots$) at
zero space and time separation.  
The locality allows us to obtain closed analytical results
at zero and finite temperatures in the limiting cases of strong and weak
interactions.  The paper is organized as follows. In section~\ref{sec:model}
we present the Hamiltonian and introduce relevant parameters of the problem.
Section~\ref{sec:strong} contains the main result of the paper, the derivation
of local correlation functions at zero and finite temperature in the
strong-coupling regime. In Section~\ref{sec:23} we use this result for the
particular case of two-body and three-body correlations. For the sake of
completeness, local correlations for the weak-coupling regime, following from
the Bogoliubov approach, are presented in Section~\ref{sec:weak}. We conclude
in Section~\ref{sec:discussion}.

\section{\label{sec:model} Interacting Bose gas in 1D}

We consider the system of $N$ bosons on a 1D ring of length $L$ in the 
thermodynamic limit $N,L\to\infty$, with a fixed density $n=N/L$. 
The particles interact via a repulsive delta-function 
interaction potential and are described  by the Hamiltonian
\begin{equation}      \label{eq:ham}
H=\frac{\hbar^2}{2M}\left[ 
\sum_{j=1}^N -\partial^2_{x_j}+2c\sum_{i<j} \delta(x_i-x_j)\right].
\end{equation}
Here $M$ is the particle mass, $x_j$ is the coordinate of the $j$-th
particle, and $c=mg/\hbar^2$, with $g>0$ being the coupling constant.
The quantity $c$ is the inverse interaction length for the 2-body problem with
the delta-function potential $g\delta(x)$. In other words, the wave function
of two particles decreases on a distance scale $1/c$ as they approach each
other.   

The Hamiltonian (\ref{eq:ham}) is diagonalized by means of the Bethe Ansatz 
\cite{LiebLiniger1963}. The many-body wave function is symmetric
with respect to permutation of particle coordinates, and 
in the domain  $0<x_1<\ldots <x_N<L$ 
the wave
function reads  
\begin{equation}
\Phi (x_1,x_2,\ldots,x_N) = \frac{1}{{\cal N}  (c)}
\sum_P a(P)\, e^{i\sum k_{P_j} x_j}. 
 \label{eq:eigfunc}
\end{equation}
In Eq.~(\ref{eq:eigfunc}) the sum runs over $N!$
permutations $P$  acting on $N$ indices of quantum numbers (quasi-momenta)
$k_j$. These quantum numbers are chosen such that
$k_1<k_2<\ldots<k_N$. They determine the amplitudes
\begin{equation}
  a(P) = \prod_{i<j} \left(\frac{ic + k_{P_i}-k_{P_j}}
    {ic - k_{P_i}+k_{P_j}}\right)^\frac{1}{2} 
  \label{eq:ap}
\end{equation}
of each term in Eq.~(\ref{eq:eigfunc}) and their quantization in a finite
system follows from the requirement of periodicity of the wave function
(\ref{eq:eigfunc}). The corresponding eigenenergy is $E=\sum k^2_j$, and
${\cal N } (c)$ in Eq.~(\ref{eq:eigfunc}) is the
normalization constant. The latter has been calculated in Refs. 
\cite{GaudinBook,Gaudin1981}. 

The key parameter of the system is the ratio of the mean interparticle
separation $1/n$ to the interaction length $1/c$:
\begin{equation}    \label{gamma}
\gamma=\frac{c}{n}=\frac{mg}{\hbar^2n}.
\end{equation}
At sufficiently low temperatures, for $\gamma\gg 1$ the gas is in the
strong-coupling regime: the many-body wave function strongly decreases at
interparticle distances much smaller than $1/n$.  
In the extreme case $\gamma=\infty$ the amplitudes (\ref{eq:ap}) are
determined by the sign of the permutation: 
\begin{equation}
\label{eq:apinfty}
a(P) = (-1)^P. 
\end{equation}
So, the wave function (\ref{eq:eigfunc}) vanishes when at least two
particle coordinates coincide. In this limit any correlation
function of the density operators can be calculated by using the theory
of free fermions, since in the domain $0<x_1<x_2<\ldots<x_N<L$ the
wave function (\ref{eq:eigfunc}) is the Slater determinant
constructed out of the plane waves with momenta $k_j$. 
In the case of large but finite $\gamma$, the calculations can be performed 
perturbatively as expounded in the next Section for the local density 
correlations.

In the opposite limit, $\gamma\ll 1$, the gas is in the weak-coupling regime.
In this case the ground state of the system is well described by the
Gross-Pitaevskii mean-field theory, and correlation functions at $T=0$ can be
found on the basis of the Bogoliubov-Popov approach (see \cite{Popov}). 
This approach also covers a significant part of the weak-coupling regime 
at finite temperatures. 

For finite temperatures, aside from $1/c$ and $1/n$, one has another length
scale, the thermal de Broglie wavelength $\Lambda=(2\pi\hbar^2/mT)^{1/2}$. The
latter is especially important at temperatures $T>T_d$, where
$T_d=\hbar^2n^2/2m$ is the temperature of quantum degeneracy.     
The relation between the three length scales determines various
finite-temperature regimes of the 1D Bose gas, which we discuss below in the
limits of strong and weak interactions.

\section{\label{sec:strong} Local correlations in the strong-coupling limit}

We now consider the strong coupling limit $\gamma\gg 1$ and
calculate the local $m$-particle correlation function
\begin{equation}
\label{eq:gm}
g_m (\gamma) = \langle \big(\Psi^\dagger (0)\big)^m \big(\Psi(0)
\big)^m\rangle,
\end{equation}
where   $\Psi^\dagger (x)$,$\Psi(x)$  are the creation and
annihilation operators of bosons. At $T=0$ the expectation value  
$\langle\ldots\rangle$ is taken with respect to
the ground state of the system, and at 
finite temperatures we assume the average in the grand-canonical ensemble.
We first present the derivation at zero temperature and then generalize the
method to the case of finite temperatures.

For $T=0$, in first quantization the correlation function
$g_m$ (\ref{eq:gm}) reads
\begin{equation}
\label{eq:gmfirst}
g_m (\gamma) = \frac{N!}{m!(N-m)!}\int dx_{m+1}\ldots dx_N \left|
\Phi^{(\gamma)}_0 (0,\ldots,0,x_{m+1},\ldots,x_N)\right|^2, 
\end{equation}
where the ground state wave function $\Phi_0^{(\gamma)}$ is given by 
by the Bethe Ansatz expression (\ref{eq:eigfunc}).
In the strong-coupling limit  the amplitudes $a(P)$ 
(\ref{eq:ap}) can be expanded in inverse powers of the interaction strength 
\begin{equation}
  a(P) = \prod_{i<j} \left(\frac{ic + k_{P_i}-k_{P_j}}
    {ic - k_{P_i}+k_{P_j}}\right)^\frac{1}{2} = 
  (-1)^P\prod_{i<j} \left(1+\frac{k_{P_i}-k_{P_j}}{ic} +\ldots\right).
  \label{eq:apstrong}
\end{equation}
The quasi-momenta are close to their ground state values at $\gamma=\infty$,
given by the uniform Fermi-Dirac distribution in the interval $[-k_F,k_F]$, 
with the Fermi momentum $k_F=\pi n$. One then sees that in fact
Eq.~(\ref{eq:apstrong}) gives the expansion in inverse powers of $\gamma$. 
For $\gamma \gg 1$, we extract the leading behavior of the wave function
$\Phi_0^{\gamma}$ at $m$ coinciding points by appropriately symmetrizing the 
amplitudes $a(P)$ for each permutation: 
\begin{equation}
  \label{eq:apsym}
  \frac{1}{m!}\sum_{p}
  a(P_{p_1},P_{p_2},\ldots,P_{p_m},P_{m+1},\ldots P_N) \simeq
   \frac{(-1)^P}{(ic)^{m(m-1)/2} }\Delta_m (k_{P_1},\ldots,k_{P_m}), 
\end{equation}
where the sum runs over $m!$ permutations $p$ of numbers $1,2,\ldots,m$, and
\begin{equation}
  \label{eq:vandermonde}
  \Delta_m (k_1,\ldots,k_m)=\prod_{i<j} (k_i-k_j)
\end{equation}
is the Vandermonde determinant. Up to the leading term in $1/c$,
the ground state wave function at $m$
coinciding points  is therefore given by 
\begin{equation}      \label{eqpr}
  \Phi^{(\gamma)}_0 (0,\ldots,0,x_{m+1},\ldots,x_N) 
  =\frac{1}{(ic)^{m(m-1)/2} }\frac{1}{{\cal N} (\infty)} 
  \sum_P (-1)^P \Delta_m ( k_{P_1},\ldots, k_{P_m}) \, 
  \exp{\left( i\sum^N_{j=m+1}k_{P_j}x_j\right)}.
\end{equation} 
This allows us to express $\Phi^{(\gamma)}_0 (0,\ldots,0,x_{m+1},\ldots,x_N)$
through spatial derivatives of the wave function of non-interacting fermions,
\begin{eqnarray*}
\Phi_0^{(\infty)}(x_1,\ldots,x_N)=\frac{1}{{\cal N} (\infty)} 
\sum_{P}(-1)^P\exp{\left(
i\sum_{j=1}^Nk_{P_j}x_j\right)}.
\end{eqnarray*} 
From Eq.~(\ref{eqpr}) we find
\begin{equation}
\Phi^{(\gamma)}_0 (0,\ldots,0,x_{m+1},\ldots,x_N) = \left.
\left(\frac{-1}{c}\right)^{m(m-1)/2}       \Delta_m ( \partial_{x_1},\ldots,
\partial_{x_m})      \Phi^{(\infty)}_0 
    (x_1,\ldots,x_m,x_{m+1},\ldots,x_N)\right|_{x_1=\ldots=x_m=0}.
\label{eq:gslead}
\end{equation}
Returning to Eq.~(\ref{eq:gmfirst}), one sees that the local correlation
function is given by derivatives of the $m$-particle correlation function of
free fermions at $x_1=\ldots=x_m=y_1=\ldots=y_m=0$:
\begin{equation}
  \label{gmder}
  g_m = \frac{1}{c^{m(m-1)}} 
  \Delta_m ( \partial_{x_1},\ldots, \partial_{x_m}) 
  \Delta_m ( \partial_{y_1},\ldots, \partial_{y_m}) 
  \langle \psi^\dagger (x_1) \ldots\psi^\dagger (x_m) 
  \psi (y_m) \ldots\psi(y_1) \rangle. 
\end{equation}
The operators 
$\psi^\dagger$, $\psi$ are now  fermionic field operators with canonical 
anticommutation relations. Using Wick's theorem we express the
expectation value of these operator as a product of one-particle Green
functions: 
\begin{equation}
  \label{eq:wick}
  \langle \psi^\dagger (x_1) \ldots\psi^\dagger (x_m) 
  \psi (y_m) \ldots\psi(y_1) \rangle = \sum_{p}(-1)^p\,
   G(x_1,y_{p_1})G(x_2,y_{p_2})\ldots G(x_m,y_{p_m}).
\end{equation}
The one-particle Green function is given by 
\begin{equation}
  \label{green}
  G(x,y)=G(x-y) = \langle \psi^\dagger (x) \psi (y) \rangle = 
  \frac{1}{L} \sum_k N^0(k)\, e^{ik(x-y)}, 
\end{equation}
where $N^0(k)$ 
is the ground state Fermi-Dirac distribution. Then, from 
Eqs.~(\ref{green}), (\ref{eq:wick}) and  (\ref{gmder}) we obtain
\begin{equation}
  \label{eq:gmfinal1}
  g_m = \frac{1}{c^{m(m-1)}} \sum_{p} \frac{(-1)^p}{L^m}
  \Delta_m(\partial_x) 
  \Delta_m(\partial_y) 
  \left.\sum_{k_1\ldots k_m} N^0_{k_1}\ldots N^0_{k_m} 
 e^{ik_1(x_1-y_{p_1})}\ldots e^{ik_m(x_m-y_{p_m})}
  \right|_{\begin{array}{c} x_1=\ldots=x_m=0\\
      y_1=\ldots=y_m=0\end{array}},
\end{equation}
where $\Delta_m(\partial_x)$ stands for 
$\Delta_m ( \partial_{x_1},\ldots, \partial_{x_m})$.
This expression can be further simplified as the Vandermonde
determinant is a totally antisymmetric function of its variables. Therefore,
reordering $y_{p_j}$ arguments cancels the factor $(-1)^p$. Using
this fact, acting with derivatives on the corresponding arguments in
the exponents, and noting that all permutations give the same result, we
arrive at the equation: 
\begin{equation}
  \label{eq:gmfinal2}
  g_m = \frac{m!}{c^{m(m-1)}} \frac{1}{L^m} 
\sum_{k_1\ldots k_m} N^0(k_1)\ldots N^0(k_m) \Delta^2_m (k_1,\ldots,k_m).
\end{equation}

The developed method was briefly outlined in Ref. \cite{Gangardt} for
3-particle local correlations. It is readily generalized to the case of finite
temperatures. In the grand canonical ensemble the expression for the
$m$-particle local correlation function $g_m$ (\ref{eq:gm}) reads:
\begin{eqnarray}
  g_m &=& \frac{1}{Z} \mbox{Tr}
  \left\{\big(\Psi^\dagger (0)\big)^m \big(\Psi(0)
  \big)^m \exp{\{-(H-\mu N)/T\}}\right\}=\nonumber\\
&=&\sum_{N=1}^\infty\frac{z^N}{N!}\sum_{\alpha} e^{-E_\alpha/T}
\frac{N!}{m!(N-m)!} 
\int dx_{m+1}\ldots dx_N \left|
    \Phi^{(\gamma)}_\alpha (0,\ldots,0,x_{m+1},\ldots,x_N)\right|^2, 
  \label{eq:gmtermdef}
\end{eqnarray}
where $Z$ is the partition function, $z=\exp(\mu/T)$ is the fugacity, $\mu$ is
the chemical potential, and the index $\alpha$ labels eigenstates of the
system. Expanding the eigenfunctions $\Phi^{(\gamma)}_\alpha$ at coinciding 
points exactly as above in the case of the ground state wave function, we 
express the leading contribution to $g_m$ through derivatives
of the finite-temperature correlation function of free fermions at 
$x_1=x_2=\ldots=x_m=y_1=\ldots=y_m=0$:
\begin{eqnarray}
  \label{eq:gmtermexpr}
  g_m 
    &=&  \frac{1}{c^{m(m-1)}} 
  \Delta_m ( \partial_x) 
  \Delta_m ( \partial_y)  
  \langle \psi^\dagger (x_1) \ldots\psi^\dagger (x_m) 
  \psi (y_m) \ldots\psi(y_1) \rangle_T 
\end{eqnarray}
The finite-temperature $m$-fermion correlation function is again calculated
with the use of Wick's theorem. This gives Eq.~(\ref{eq:wick}) in which
$G(x_i,y_{P_i})$ are now finite-temperature Green functions for free fermions.
We then proceed in the same way as at $T=0$ and arrive at
Eq.~(\ref{eq:gmfinal2}), with $N^0 (k)$ replaced by the Fermi-Dirac
distribution at finite temperatures:  
\begin{equation}
\label{eq:nk}
N(k)=\frac{z \exp{(-\Lambda^2k^2/4\pi)}}{1+z \exp{(-\Lambda^2k^2/4\pi)}}.
\end{equation}

At temperatures $T\gg T_d$ the
characteristic momentum of particles is the thermal momentum $k_T\sim
1/\Lambda$. Therefore, the small parameter for the expansion of the amplitudes
$a(P)$ in Eq.~(\ref{eq:apstrong}) is $1/\Lambda c$. Thus, one must satisfy 
the inequality $\Lambda c\gg 1$, which requires temperatures
\begin{equation}    \label{ineqT}
T\ll\gamma^2T_d.
\end{equation}

From this point on we consider together the cases of zero and finite
temperature. In the thermodynamic limit, expressing the momenta in terms of
the size of the Fermi zone, $k=2 k_F x=2 \pi n x$, 
we have 
\begin{equation}
  \label{eq:gmtemp}
  \frac{g_m}{n^m} = 
  \left(\frac{2\pi}{\gamma}\right)^{m(m-1)}
  m!\; I_m\left(\Lambda n,z\right),  
\end{equation}
where the  $m$-fold integral $I_m(\Lambda n,z)$ depends on the ratio of the de 
Broglie wavelength to the mean interparticle separation and on the fugacity
$z$ (or chemical potential), and is given by  
\begin{equation}
  \label{eq:integral}
  I_m\left(\Lambda n,z\right) = \int_{-\infty}^{+\infty}
  dx_1\ldots dx_m N(x_1)\ldots N(x_m) \Delta_m^2 (x_1,\ldots,x_m), 
\end{equation}
where $N(x)$ is given by Eq.~(\ref{eq:nk}) with $k=2\pi nx$.
In order to keep the density constant one must fix 
$z$ as a function of $\Lambda n$ by the normalization condition   
\begin{equation}
  \label{eq:density}
  \int_{-\infty}^{+\infty}  N(x)dx=1.
\end{equation}
Under this condition the integral in Eq.~(\ref{eq:integral}) 
becomes a function of $\Lambda n$ only.
For calculating this function we employ the method of orthogonal polynomials 
used in Random Matrix theory \cite{MehtaRandMatr}. Consider a set of
polynomials $P_j(x)=x^j+\ldots$ for  $j=0,1,2,\ldots$,
orthogonal with the weight $N(x)$ on an infinite interval: 
\begin{equation}
  \label{eq:orthpoly}
  \int_{-\infty}^{+\infty} N(x) P_i (x) P_j (x)  dx=h_i\delta_{ij}.
\end{equation}
Then the value of the integral  (\ref{eq:integral}) is expressed through the
normalization coefficients $h_j$ as follows:
\begin{equation}
  \label{eq:integral1}
  I_m\left(\Lambda n,z\right) = m!
  h_0\ldots h_{m-1}.
\end{equation}

At $T=0$ we have $\Lambda n=\infty$. In this
limit the distribution function $N(x)$ is uniform in the interval $-1/2<x<1/2$,
and is zero otherwise. Polynomials $P_j(x)$ can be expressed in terms of
Jacobi  polynomials \cite{AbramowitzStegunChap22} and we obtain  
\begin{equation}
  \label{eq:ilegendre}
  I_m (\Lambda n) = m! \prod_{j=0}^{m-1} 
  \frac{\left[\Gamma(j+1)\right]^4}{\Gamma(2j+1)\Gamma(2j+2)},
  \;\;\;\;\Lambda n =\infty .
\end{equation}
{}From Eq.~(\ref{eq:gmtemp}) we then obtain the local correlation function 
\begin{equation}    \label{zerogm}
\frac{g_m(\gamma)}{n^m}=A_m\left(\frac{\pi}{\gamma}\right)^{m(m-1)},
\end{equation}
where the coefficients $A_m$ satisfy the recurrence relation
\begin{equation}      \label{recurrent0}
\frac{A_{m+1}}{A_m}=\pi\frac{2m+1}{2^{2m+2}}\left[\frac{\Gamma(m+2)}
{\Gamma(m+3/2)}\right]^2,\;\;\;\;\;\;\;A_1=1 .
\end{equation}

Although general expressions for the normalization coefficients $h_j$ in 
Eq.~(\ref{eq:orthpoly}) are not known, for sufficiently small $m$
these coefficients can be calculated straightforwardly as they are related to 
the moments of the distribution function $N(x)$. For example, one finds   
\begin{equation}
  \label{eq:moments}
  h_0=1,\,\,\,\,h_1 =<x^2>,\;\;\;\;h_2=<x^4>-<x^2>^2,\;\;\;\;
<x^{2j}>=\int_{-\infty}^{+\infty} x^{2j} N(x) dx.
\end{equation}

We finally consider high temperatures $T\gg T_d$ that still satisfy
Eq.~(\ref{ineqT}). In this case the fugacity is $z=\Lambda n$, and  the
distribution function is Gaussian: $N(x)=z\exp{(-\pi\Lambda^2 n^2 x^2)}$. The
polynomials $P_j (x)$ are Hermite polynomials, and we find  
\begin{equation}
  \label{eq:igauss}
I_m(\Lambda n) =\frac{m!}{{(\sqrt{2\pi}\Lambda n)^{m(m-1)}}}
  \prod_{j=0}^{m-1} \Gamma (1+j) ,
  \;\;\;\;1/c\,\ll\Lambda\ll 1/n.
\end{equation}
Then Eq.~(\ref{eq:gmtemp}) for the local correlation function takes the form
\begin{equation}      \label{Tgm}
\frac{g_m}{n^m}=B_m\left(\frac{\sqrt{2\pi}}{\Lambda
c}\right)^{m(m-1)},\,\,\,\,\,\,\,\,\,\,1/c\,\ll\Lambda\ll 1/n,   
\end{equation}
where the recurrence relation for the coefficients $B_m$ reads 
$B_{m+1}=(m+1)\Gamma(m+2)\,B_m$, and $B_1=1$. 

\section{\label{sec:23} Three-body and two-body correlations}

In this Section we use the general results of Section~\ref{sec:strong} for the
particular case of three-body local correlations in the
strong-coupling limit. Two-body correlations have been discussed in Ref.
\cite{Karen} and are expounded here for completeness.  

At temperatures $T\ll T_d$ we have $\Lambda n\gg 1$, and the local correlation
functions $g_2$ and $g_3$ are close to their zero temperature
values. Thus, the system remains in the Tonks-Girardeau
regime. For calculating $g_2$ and $g_3$ we use Eqs.~(\ref{eq:gmtemp}) and 
(\ref{eq:integral1}), with normalization coefficients $h_1$ and $h_2$ 
following from Eq.~(\ref{eq:moments}).
The quantities $< x^2 >$ and $< x^4 >$ at finite $T$ are obtained on the
basis of the Sommerfeld expansion:  
\begin{equation}
\label{eq:sommerfeld}
<x^{2j}>_T-<x^{2j}>_0 \simeq \frac{2}{3}\frac{2j}{2^{2j}}
\frac{1}{(\Lambda n)^4}=\frac{1}{24\pi^2}\frac{2j}{2^{2j}}\left(\frac{T}{T_d}\right)^2.
\end{equation}
This gives the following expressions:
\begin{eqnarray} \label{eq:g2strongsmallt}
\frac{g_2}{n^2}&=&
\frac{4}{3}\left(\frac{\pi}{\gamma}\right)^2
\left[1+\frac{4}{(\Lambda n)^4}\right]
=\frac{4}{3}\left(\frac{\pi}{\gamma}\right)^2
\left[1+\frac{1}{4\pi^2}\left(\frac{T}{T_d}\right)^2\right]
\\
\label{eq:g3strongsmallt}
\frac{g_3}{n^3}&=&
\frac{16}{15}\left(\frac{\pi}{\gamma}\right)^6
\left[1+\frac{28}{(\Lambda n)^4}\right]
=\frac{16}{15}\left(\frac{\pi}{\gamma}\right)^6
\left[1+\frac{7}{4\pi^2}\left(\frac{T}{T_d}\right)^2\right]
\end{eqnarray}

The results of Eqs.~(\ref{eq:g2strongsmallt}) and (\ref{eq:g3strongsmallt}) 
have a clear physical explanation. For $T\ll T_d$ the characteristic 
momentum of particles is of the order of $k_F=\pi n$. Fermionic 
correlations are present at interparticle distances $x\agt 1/c$, since for 
smaller distances the correlations do not change. Therefore, 
$g_2$ is nothing else than the pair correlation function for free fermions 
at a distance $x\sim 1/c$ \cite{Karen}. This function is of the order of 
$(k_F/c)^2\sim (\pi/\gamma)^2$, which agrees with Eq.~(\ref{eq:g2strongsmallt}). 
With the same arguments, one finds that the 3-body correlation function is 
$g_3\sim (k_F/c)^2\sim (\pi/\gamma)^6$, which coincides with the result of
Eq.~(\ref{eq:g3strongsmallt}).

For temperatures in the interval $T_d\ll T\ll \gamma^2T_d$,
where $1/c\ll\Lambda\ll 1/n$, we use Eq.~(\ref{Tgm}) directly and
obtain: 
\begin{eqnarray}
\label{eq:g2stronglarget}
\frac{g_2}{n^2}&=&\frac{8\pi}{(\Lambda c)^2}
=\frac{2}{\gamma^2}\frac{T}{T_d}
\\
\label{eq:g3stronglarget}
\frac{g_3}{n^3}&=&\frac{576\pi^3}{(\Lambda c)^6}=
\frac{9}{\gamma^6}\left(\frac{T}{T_d}\right)^3 .
\end{eqnarray}
This regime can be called the regime of ``high-temperature fermionization"
\cite{Karen}. The gas is no longer degenerate, but due to strong interaction
between particles the correlations still have a fermionic character. With
regard to local correlations, the main difference from the low temperature
case is related to the value of the characteristic momentum of particles. It
is now of the order of $1/\Lambda$, instead of $k_F$ at $T\ll T_d$.
Accordingly, the two-body and three-body local correlation functions are
$g_2\sim (1/\Lambda c)^2$ and $g_3\sim (1/\Lambda c)^6$, as described by 
Eqs.~(\ref{eq:g2stronglarget}) and (\ref{eq:g3stronglarget}).   

\section{\label{sec:weak}Weak-coupling limit}

In the weak-coupling limit, where $\gamma\ll 1$, one can rely on the
mean-field approach. The mean-field interaction energy per particle
is proportional to $ng$ and it is reasonable to introduce the correlation
length $l_c=\hbar/\sqrt{mng}$. At temperatures $T\ll T_d$, over a wide range of
parameters the correlation length $l_c\ll l_{\phi}$, where $l_{\phi}$ is the
phase coherence length. Then the equilibrium state is a quasicondensate,
that is the state in which density fluctuations are suppressed but the phase 
still fluctuates \cite{Mermin1966}. A review of earlier studies of 1D Bose
gases in the weak-coupling limit may be found in Ref. \cite{Popov}. 
The zero temperature phase coherence length is always larger than $l_c$,
and at finite $T$ one has $l_{\phi}=\hbar^2n/mT$. Therefore, the condition
$l_c\ll l_{\phi}$ can be rewritten as
\begin{equation}     \label{quasi}
\frac{T}{T_d}\ll \sqrt{\gamma}.
\end{equation}
One then sees that the quasicondensate regime requires sufficiently low
temperatures or relatively large interaction between particles.

Due to the presence of phase coherence on a long distance scale, local
correlation functions in the quasicondensate regime are close to $n^2$.
Deviations from this value can be calculated straightforwardly by using the
Bogoliubov approximation \cite{Mora}. For $g_2$ the results are known (see
\cite{Castin,Karen} and references therein), and $g_3$ at $T=0$ is given in
Ref. \cite{Gangardt}. Here we complete the picture and present general
results for $g_m$. They are obtained in the same way as $g_2$ in Ref.
\cite{Karen}. 

We represent the bosonic field operator as a sum of a macroscopic
component $\Psi _{0}$ and a small component $\Psi ^{\prime }$ describing
finite-momentum excitations. To be more precise, the component $\Psi _{0}$
contains the contribution of excitations with momenta 
$k\lesssim k_{0}\ll l_{c}^{-1}$,
whereas $\Psi ^{\prime }$ includes the contribution of larger $k$.
At the same time, the momentum $k_{0}$ is chosen such that most of
the particles are contained in the part $\Psi _{0}$. This picture
is along the lines of Ref. \cite{Popov}, and the momentum $k_{0}$
drops out of the answer as the main contribution of the excitation
part $\Psi ^{\prime }$ to local correlation functions is provided
by excitations with $k\sim l_{c}^{-1}$. Confining ourselves to the terms 
quadratic in $\Psi^{\prime}$ and taking then into account that 
$|\Psi_0|^{2m}=n^m-mn^{m-1}\langle\Psi^{\prime \dagger}\Psi^{\prime}\rangle$, 
the $m$-particle local correlation function is given by
\begin{equation}     \label{GPgm}
g_m=\langle \Big(\Psi_0^*+\Psi^{\prime \dagger}\Big)^m
\Big(\Psi_0+\Psi^{\prime}\Big)^m\rangle
=n^m\left[
1-\frac{m(m-1)}{n}\Big(\langle\Psi^{\prime \dagger}\Psi^{\prime}\rangle+
\langle\Psi^{\prime}\Psi^{\prime}\rangle\Big)\right].
\end{equation} 

The normal and anomalous averages, $\langle \Psi ^{\prime }{}^{\dagger }\Psi
^{\prime }\rangle $ and $\langle \Psi ^{\prime }\Psi ^{\prime }\rangle $, can
be calculated by using the same Bogoliubov transformation for $\Psi ^{\prime }$
as in the 3D case: 
\begin{equation}     \label{Bog}
\Psi^{\prime}=\sum_k \left(u_k a_k e^{ikx}-v_k a_k^{\dagger}e^{-ikx}\right),
\end{equation}
where $a_k$ and $a_k^{\dagger}$ are annihilation and creation operators of
elementary excitations, $u_k,v_k=(\varepsilon_k\pm
E_k)/2\sqrt{\varepsilon_kE_k}$, $\varepsilon_k=\sqrt{E_k^2+2ngE_k}$ is the
Bogoliubov excitation energy, and $E_k=\hbar^2k^2/2m$. This immediately gives
\begin{equation}
\frac{g_m}{n^m}=1+m(m-1)\int _{-\infty }^{\infty }\frac{dk}{2\pi
n}\left[\frac{E_{k}}{\varepsilon _{k}}(1+\tilde N_{k})-1\right],
\label{eq:g2weak}
\end{equation}  
with $\tilde N_{k}=[\exp{(\varepsilon_k/T)}-1]^{-1}$ being the equilibrium
occupation numbers for the Bogoliubov  excitations.

The integral term in Eq.~(\ref{eq:g2weak}) contains the contribution
of both vacuum and thermal fluctuations. The former is determined
by excitations with $k\sim l_{c}^{-1}$, and at $T=0$ we find
\begin{equation}
  \label{eq:gmsmallres}
  \frac{g_m}{n^m} = 1-\frac{m(m-1)}{\pi} \sqrt{\gamma}.
\end{equation}
At temperatures $T\ll ng\sim\gamma T_d$, thermal fluctuations provide only a
small correction $(\pi\sqrt{\gamma}/24)(T/\gamma T_d)^2$ to the result of
Eq.~(\ref{eq:gmsmallres}).

For temperatures $T\gg ng\sim \gamma T_d$ (but still $T\ll
\sqrt{\gamma}T_d$), the contribution of thermal fluctuations to the integral 
term in Eq.~(\ref{eq:g2weak}) is the most important. It comes from excitation energies
$\varepsilon_k\sim ng$ and, hence, one may put $\tilde N(k)=T/\varepsilon_k$
in the integrand. This yields
\begin{equation}
\label{tempcorrtgggn}
\frac{g_m}{n^m} =
  1+\frac{m(m-1)}{4}\frac{T}{\sqrt{\gamma}T_d}.
\end{equation}

One clearly sees from Eq.~(\ref{tempcorrtgggn}) that $g_m$ increases with
decreasing interaction strength or increasing temperature. The ratio
$T/T_d\sqrt{\gamma}$ is simply $l_c/l_{\phi}$. For $l_c\sim
l_{\phi}$ or $T\sim T_d\sqrt{\gamma}$, the gas leaves the quasicondensate
regime and Eq.~(\ref{tempcorrtgggn}) is no longer valid. At a temperature and 
interaction strength satisfying the condition $T\gg T_d\sqrt{\gamma}$ the gas
is in the decoherent regime where the interaction between particles is much
less important. Note that for a very small interaction strength this regime is
present even at temperatures far below $T_d$. The cross-over from the
quasicondensate to the decoherent regime was discussed in Refs.
\cite{Castin,Karen} on the basis of calculations for $g_2$. In the decoherent
regime the local correlation function is close to the result
for an ideal Bose gas following from Wick's theorem: $g_m/n^m=m!$. The
corrections to this value can be calculated perturbatively \cite{Karen}. The
lower is $T$, the smaller is $\gamma$ at which the gas enters the decoherent
regime. For $T=0$ this transition is discontinuous and occurs at zero
interaction strength.

\section{\label{sec:discussion} Concluding remarks}

From a general point of view, local correlations of the 1D system are less
universal than the long-wave behavior. Local correlation functions depend on a 
particular model used for calculations.
Our approach leads to physically transparent analytical results for $g_m$ of
the 1D Bose gas in the Lieb-Liniger model. 

The use of this model for trapped atomic
gases in the 1D regime is justified by the short-range character of
interatomic interaction: the characteristic radius of interactions is much
smaller than any length-scale of the Lieb-Liniger model. 
For an atomic gas in an infinitely long cylindrical trap, harmonically
confined with frequency $\omega_0$ in the transverse direction to zero-point
oscillations, the coupling constant $g$ is expressed through the 3D scattering
length $a$ and reads \cite{Olshanii98} $g=2\hbar^2 a/M l_0^2$, where
$l_0=\sqrt{\hbar/M\omega_0}$ is the amplitude of
zero point oscillations. Then, to describe the system by the 1D Hamiltonian
(\ref{eq:ham}), it is sufficient to satisfy the inequalities 
$l_0<< 1/n, \Lambda_T$ \cite{Gangardt,Karen}.

The reduction of $g_3$ in the strong-coupling limit, which follows from our
results, is important in two aspects. First, it indicates a possibility of
achieving this limit at a high gas density (large number of particles), since
the rate of decay due to 3-body recombination is proportional to $g_3$ and
will be strongly suppressed. Second, the measurement of 3-body losses of
particles can be used for identifying the strong-coupling
Tonks-Girardeau regime and the regime of high-temperature fermionization.
The identification of these regimes can also be provided by the measurement of
photoassociation in pair atomic collisions as the rate of this process is
proportional to the two-particle local correlation function $g_2$. In this
respect, it is worth mentioning that recent Monte Carlo calculations of
$g_2$ at $T=0$ for the number of particles as low as 100 \cite{Giorgini} agree
with our results obtained in the thermodynamic limit. 

\begin{acknowledgments}

We are grateful to Lincoln Carr for useful comments and we acknowledge the
hospitality of the European Centre for Theoretical Studies in Nuclear Physics
and Related areas (ECT) during the 2002 BEC Summer Program, where part of
the present work was done. This work was financially supported by the 
French Minist\`ere des Affaires Etrang\`eres, by 
the Dutch Foundations NWO and FOM, by INTAS, and by the Russian Foundation 
for Fundamental Research. Laboratoire Kastler Brossel is a research unit of 
Universit\'e Pierre et Marie Curie and Ecole Normale Sup\'erieure, 
associated with CNRS (UMR 8552).

\end{acknowledgments}

\end{document}